\documentclass[preprint,showpacs,preprintnumbers,amsmath,amssymb]{revtex4}
\usepackage{amsmath,amssymb,graphics,epsfig,subfigure}
\usepackage{cases}
\usepackage{color}
\usepackage[colorlinks,
            linkcolor=blue,
            anchorcolor=blue,
            citecolor=green
            ]{hyperref}
\usepackage[colorlinks,linkcolor=blue,anchorcolor=blue,citecolor=green]{hyperref}
\allowdisplaybreaks

\begin{document}

\allowdisplaybreaks

\title{Generalized Approach for the Perturbative Dynamical Braneworld in $D$ Dimensions}

\author{Zi-Chao Lin$^a$$^b$\footnote{linzch22@hust.edu.cn},
        Hao Yu$^a$\footnote{yuhaocd@cqu.edu.cn},
        Yungui Gong$^c$$^b$\footnote{gongyungui@nbu.edu.cn (corresponding author)}}

\affiliation{$^{a}$College of Physics, Chongqing University, Chongqing 401331, China\\
$^{b}$School of Physics, Huazhong University of Science and Technology, Wuhan, Hubei 430074, China\\
$^{c}$Department of Physics, School of Physical Science and Technology, Ningbo University, Ningbo, Zhejiang 315211, China}

\begin{abstract}
{In this paper, we propose an approach to derive the brane cosmology in the $D$-dimensional braneworld model. We generalize the ``bulk-based'' approach by treating the 4-brane as a small perturbation to the $D$-dimensional spherically symmetric spacetime. The linear corrections from a static 4-brane to the metric are derived from the linearized perturbation equations, while the nonlinear corrections are found by a parameterization of the perturbed metric solution. We use a time-dependent generalization to give the nonlinearly perturbed metric solution for the dynamical braneworld model, and analyze the stability of the model under the motion of the 4-brane. Through the fine tuning, we can recover the Friedmann equations for the universe with and without an effective cosmological constant. More importantly, the de Sitter expansion of the universe can be reproduced.}
\end{abstract}

%\keywords{Black holes, critical phenomena, phase diagram}

%\pacs{04.50.Kd, 04.50.+h}

\maketitle

\section{Introduction}

The idea that there exist extra (spatial) dimensions is compelling. It has sparked numerous discussions regarding the nature and the structure of extra dimensions. In the early higher-dimensional theories, it was commonly believed that these extra dimensions are compact and of small size~\cite{Kaluza1,Klein1,Klein2,ADD1,ADD2,Randall1}. An opposing viewpoint, initially proposed by Rubakov and Shaposhnikov~\cite{Rubakov1,Rubakov2}, emerged in the model developed by Randall and Sundrum (i.e., the RS-2 model)~\cite{Randall2}, in which the extra dimensions could be large and infinite. As a braneworld model, the universe in the RS-2 (and many other) model is supposed to be a 4-brane embedded in a static anti-de Sitter (AdS) bulk. All the particles in the standard model are confined on the 4-brane and the linearized gravity on the 4-brane is proved to be Einstein gravity~\cite{Randall2,Lykken:1999nb,Garriga1,Giddings1}. However, it is challenging to directly apply the RS-like model to brane cosmology. In fact, if the embedded 4-brane has a de Sitter (dS) expansion, the bulk spacetime is generally time-dependent. While in the model the fine tuning between the bulk cosmological constant and the energy density and pressure of the 4-brane could keep the bulk spacetime static, it will result in a vanishing effective brane cosmological constant. Thus, the induced metric on the 4-brane inevitably becomes flat Minkowski, which fails to describe the dS expansion.

To find the cosmological brane solutions for the RS spacetime, a ``brane-based'' approach was proposed in Refs.~\cite{Kaloper:1999sm,Chung:1999zs,Csaki:1999jh,Cline:1999ts,Shiromizu:1999wj,Maartens:2000fg}. Notably, it was the Bin\'{e}truy-Deffayet-Langlois (BDL) model that exactly solved the cosmological equations on a single 4-brane embedded in a 5-dimensional bulk governed by a negative bulk cosmological constant~\cite{Binetruy1,Binetruy2}. In the BDL model, the 4-brane is fixed in the bulk in order to get Gaussian normal coordinates, with which the extra diagonal metric component becomes normal to the 4-brane. Unlike the RS-like single-wall model, the bulk spacetime in the BDL model is supposed to be time-dependent to induce the cosmological expansion of the 4-brane. The evolution of the bulk spacetime is driven by the matter on the 4-brane through the singularity part of the field equations. So the background spacetime built in the ``brane-based'' approach could be understood as a time-dependent generalization of the static bulk spacetime with time-dependent corrections sourced by the energy density and pressure of the 4-brane (see also~\cite{Cline:1999ts}).

An equivalent description of the ``brane-based'' approach can be achieved by applying an explicit coordinate transformation to the Gaussian normal coordinates~\cite{Mukohyama:1999wi,Bowcock:2000cq}. Instead of keeping static in the bulk, the 4-brane in the alternative ``bulk-based'' approach could be dynamical due to the energy density and pressure of the 4-brane~\cite{Kraus:1999it,Ida:1999ui,Chamblin:1999ya,Chamblin:1999ea,Gubser:1999vj,Hawking:2000kj,Anchordoqui:2000du}. The time-dependent cosmological equations on such a dynamical 4-brane are usually derived from the Israel joint condition~\cite{Israel1}. In this approach, the static bulk spacetime structure is more transparent because it exhibits maximal symmetry under the motion of the 4-brane. Taking advantage of this, several researches, including brane cosmology~\cite{Collins:2000yb,Banerjee:2018qey,Csaki:2000dm}, AdS/CFT~\cite{Garriga:1999bq,Savonije:2001nd,Cai:2001ja}, shortcuts~\cite{Caldwell1,Wang2002,Visinelli1,Yu1,Lin1,Lin2,Lin:2023zna}, etc., have been conducted from the ``bulk-based'' perspective.

Note that neither of the two approaches considers both the back-reaction and the dynamics of the 4-brane. In this paper, we aim to provide a generalized version of the ``bulk-based'' approach by considering the back-reaction from a dynamical 4-brane to the bulk spacetime. The contributions from the 4-brane are treated as small perturbations to the bulk, enabling us to analytically calculate the linear corrections to the metric. The nonlinear corrections are introduced through the parameterization of the perturbed metric to stabilize the dynamical braneworld model. We also study the cosmological equations on the 4-brane and prove that {there exists a dS expansion on the 4-brane.}

The paper is arranged as follows. In Sec.~\ref{sec2}, we derive the Einstein equations for a $D$-dimensional spherically symmetric spacetime with a bulk cosmological constant and a 4-brane fixed in the bulk. Then, treating the static 4-brane as a small perturbation, we obtain a nonlinearly perturbed metric solution by a specific parameterization. By introducing a time-dependent generalization of the parameterization, we establish a perturbative dynamical braneworld model in Sec.~\ref{sec3}. {We also study the cosmological equations on the 4-brane, and show the normal expansions of the universe.} Finally, our conclusion and discussion are given in Sec.~\ref{sec4}.

\section{Static 4-brane}\label{sec2}

\subsection{Field equations for the $D$-dimensional bulk}

We start by considering a $D(=4+d)$-dimensional spacetime with a bulk cosmological constant $\Lambda$ and an embedded 4-brane. The gravity is described by the $D$-dimensional general relativity with its action written as
\begin{equation}
	S_{g}^{~}=M^{D-2}_{~}\int\text{d}^{D}_{~}x\sqrt{-g}\,(R^{(D)}_{~}-2\Lambda).
\end{equation}
Here, $M$ is the $D$-dimensional fundamental Planck scale. With the existence of the 4-brane, the $D$-dimensional spacetime is not static in general. The metric could be given by
\begin{equation}\label{bm1}
	\text{d}s^{2}_{D}=A(R,T)\text{d}R^{2}_{~}+B(R,T)\text{d}\Omega^{2}_{d-1}+C(R,T)\text{d}\Sigma^{2}_{4}\,,
\end{equation}
where
\begin{equation}
	\text{d}\Omega^{2}_{d-1}=\tilde{g}_{mn}^{~}\text{d}y^{m}_{~}\text{d}y^{n}_{~}
\end{equation}
is the line element of a $(d-1)$-dimensional subspace, $T$ is the bulk time, and $(R,y^{m}_{~})$ are the coordinates of the extra space. We use the line element
\begin{equation}
	\text{d}\Sigma^{2}_{4}=\hat{g}_{\mu\nu}^{~}\text{d}x^{\mu}_{~}\text{d}x^{\nu}_{~}
\end{equation}
to describe a static and maximally symmetric 4-dimensional submanifold, where $x^{\mu}_{~}=(T,r,\theta,\phi)$ represents the 4-dimensional coordinates spanning on it. The $TT$ component of $\hat{g}_{\mu\nu}^{~}$ can be set to $\hat{g}_{TT}^{~}=-1$ to simplify the following calculations based on a coordinate transformation on $T$.

Without the loss of generality, we assume that the 4-brane can move in the bulk. However, as we shall see in Sec.~\ref{sec3}, the evolution of the bulk metric is governed by both the brane's motion and the matter confined on it. The contributions from the two sources are coupled, making it challenging to obtain the metric solutions directly. As a result, we would like to consider a static 4-brane at first. Under the metric assumption~\eqref{bm1}, the 4-brane is embedded through the condition
\begin{equation}\label{bp1}
	R=R_{0}^{~}\,,~~~~y^{m}_{~}=y_{0}^{m}\,,
\end{equation}
where $R_{0}^{~}$ and $y_{0}^{m}$ denote a fixed position in the bulk. With this condition, the 4-dimensional fields confined on the brane will couple to the following induced metric:
\begin{equation}
	g^{(4)}_{\mu\nu}\equiv g_{\mu\nu}^{~}(x^{\mu}_{~},R=R_{0}^{~},y^{m}_{~}=y_{0}^{m}).
\end{equation}
The energy-momentum tensor of the matter confined on the 4-brane is
\begin{equation}\label{emt1}
	T_{MN}^{~}\!=\!\big[(\rho+p)u_{M}^{~}u_{N}^{~}+p\,h_{MN}^{~}\big]\delta(R-R_{0}^{~})\delta^{(d-1)}_{~}(y^{m}_{~}-y_{0}^{m}),
\end{equation}
where $\rho$ and $p$ are respectively the energy density and pressure of the matter, $u^{M}_{~}=(1/\sqrt{C},0,\ldots,0)$ is the unit velocity vector of the comoving observer on the 4-brane, and $h_{MN}^{~}=g^{(4)}_{\mu\nu}\delta^{\mu}_{M}\delta^{\nu}_{N}$ is the projection tensor for the 4-brane. The singular behavior of the energy-momentum tensor means that it will contribute a boundary condition to the bulk metric on the location of the 4-brane. When taking it into account, the Einstein equations $\sqrt{-g}\,G_{MN}^{~}=\frac{1}{M_{~}^{D-2}}\sqrt{-g^{(4)}_{~}}\,T_{MN}^{~}-\Lambda\sqrt{-g}\,g_{MN}^{~}$ become
\begin{subequations}\label{fe1}
	\begin{eqnarray}
	    \hat{G}_{ij}^{(4)}
	    -\frac{C}{2B}\hat{g}^{~}_{ij}\tilde{R}^{(d-1)}_{~}
	     &\!=\!& 
	    \frac{1}{M^{D-2}_{~}}\frac{\sqrt{-g^{(4)}_{~}} }{\sqrt{-g} }T_{ij}^{~}
	    -C\Lambda \hat{g}_{ij}^{~}
	    +\bigg[\frac{\ddot{C}}{C}-\frac{3}{2}\frac{C''}{A}+\frac{3}{4}\frac{A'C'}{A^{2}_{~}}-\frac{3}{4}\frac{\dot{C}^2_{~}}{C^2_{~}}+\frac{1}{4}\frac{\dot{A} \dot{C}}{A C}\nonumber\\
	    &\!~\!&
	    +\frac{1}{2}\frac{\ddot{A}}{A}-\frac{1}{4}\frac{\dot{A}^2_{~}}{A^2_{~}}+\frac{d-1}{4}\frac{\dot{A} \dot{B}}{A B}+\frac{d-1}{4}\frac{A' B'C}{A^2_{~} B}+\frac{d-1}{2}\frac{\ddot{B}}{B}-\frac{d-1}{2}\frac{  B''C}{A B}\nonumber\\
	    &\!~\!&
	    -\frac{(d-4)(d-1)}{8}\frac{B'^2_{~}C}{A B^2_{~}}+\frac{d-1}{4}\frac{\dot{B} \dot{C}}{B C}-\frac{3(d-1)}{4}\frac{B' C'}{A B}\nonumber\\
	    &\!~\!&
	    +\frac{(d-4)(d-1)}{8}\frac{\dot{B}^2_{~}}{B^2_{~}}\bigg]\hat{g}_{ij}^{~},\\
	     \hat{G}_{TT}^{(4)}+\frac{C }{2 B}\tilde{R}^{(d-1)}_{~}
	      &\!=\!& 
	     \frac{1}{M^{D-2}_{~}}\frac{\sqrt{-g^{(4)}_{~}} }{\sqrt{-g} }T_{TT}^{~}+C\Lambda
	     +\frac{3}{2}\frac{C''}{A}-\frac{3}{4}\frac{\dot{C}^{2}_{~}}{C^{2}_{~}}-\frac{3}{4}\frac{A'C'}{A^{2}_{~}}-\frac{3}{4}\frac{\dot{A}\dot{C}}{AC}\nonumber\\
	    &\!~\!&
	    -\frac{d-1}{4}\frac{\dot{A}\dot{B}}{AB}-\frac{(d-2)(d-1)}{8}\frac{\dot{B}^{2}_{~}}{B^{2}_{~}}-\frac{d-1}{4}\frac{A'B'C}{A^{2}_{~}B}+\frac{d-1}{2}\frac{B''C}{AB}\nonumber\\
	    &\!~\!&
	    +\frac{(d-4)(d-1)}{8}\frac{B'^{2}_{~}C}{AB^{2}_{~}}-\frac{3(d-1)}{4}\frac{\dot{B}\dot{C}}{BC}+\frac{3(d-1)}{4}\frac{B'C'}{AB},\\
	     \frac{A}{2B}\tilde{R}^{(d-1)}_{~}+\frac{A}{2 C}\hat{R}^{(4)}_{~}
	      &\!=\!& 
	      A\Lambda-\frac{3}{2}\frac{A\ddot{C}}{C^{2}_{~}}+\frac{3}{2}\frac{C'^{2}_{~}}{C^{2}_{~}}+\frac{3}{4}\frac{A\dot{C}^{2}_{~}}{C^{3}_{~}}-\frac{d-1}{2}\frac{A\ddot{B}}{BC}+\frac{(d-2)(d-1)}{8}\frac{B'^{2}_{~}}{B^{2}_{~}}\nonumber\\
	    &\!~\!&
	    -\frac{(d-4)(d-1)}{8}\frac{A\dot{B}^{2}_{~}}{B^{2}_{~}C}-\frac{d-1}{2}\frac{A\dot{B}\dot{C}}{BC^{2}_{~}}+(d-1)\frac{B'C'}{BC},\\
	     \tilde{G}^{(d-1)}_{mn}-\frac{B}{2C}\tilde{g}^{~}_{mn}\hat{R}^{(4)}_{~}
	      &\!=\!& 
	      -B\Lambda\tilde{g}_{mn}^{~}+\bigg[\frac{3}{2}\frac{B\ddot{C}}{C^{2}_{~}}-2\frac{BC''}{AC}-\frac{1}{2}\frac{BC'^{2}_{~}}{AC^{2}_{~}}-\frac{3}{4}\frac{B\dot{C}^{2}_{~}}{C^{3}_{~}}+\frac{A'BC'}{A^{2}_{~}C}+\frac{1}{2}\frac{\ddot{A}B}{AC}\nonumber\\
	    &\!~\!&
	    +\frac{1}{2}\frac{\dot{A}B\dot{C}}{AC^{2}_{~}}+\frac{d-2}{4}\frac{\dot{A}\dot{B}}{AC}-\frac{1}{4}\frac{\dot{A}^{2}_{~}B}{A^{2}_{~}C}+\frac{d-2}{4}\frac{A'B'}{A^{2}_{~}}+\frac{d-2}{2}\frac{\ddot{B}}{C}-\frac{d-2}{2}\frac{B''}{A}\nonumber\\
	    &\!~\!&
	    -\frac{(d-5)(d-2)}{8}\frac{B'^{2}_{~}}{AB}+\frac{d-2}{2}\frac{\dot{B}\dot{C}}{C^{2}_{~}}-(d-2)\frac{B'C'}{AC}
	     \bigg]\tilde{g}_{mn}^{~},\\
	     \frac{3}{2}\frac{\dot{C}C'}{C^{2}_{~}}-\frac{3}{2}\frac{\dot{C}'}{C}
	       &\!=\!& 
	      \frac{d-1}{2}\frac{\dot{B}'}{B}-\frac{d-1}{4}\frac{\dot{B}B'}{B^{2}_{~}}-\frac{3}{4}\frac{\dot{A}C'}{AC}-\frac{d-1}{4}\frac{\dot{B}C'}{BC}-\frac{d-1}{4}\frac{\dot{A}B'}{AB},
    \end{eqnarray}
\end{subequations}
where primes denote derivatives with respect to $R$, $\tilde{R}^{(d-1)}_{~}$ is the curvature scalar of the $(d-1)$-dimensional subspace constructed by $\tilde{g}_{mn}^{~}$, $\hat{R}^{(4)}_{~}$ is the curvature scalar of the 4-dimensional submanifold associated with $\hat{g}_{\mu\nu}^{~}$, and $\tilde{G}_{mn}^{(d-1)}$ and $\hat{G}^{(4)}_{\mu\nu}$ are induced Einstein tensors related to $\tilde{R}^{(d-1)}_{~}$ and $\hat{R}^{(4)}_{~}$, respectively. However, it is hard to solve these field equations directly due to the numerous parameters involved. To decouple the effect of the 4-brane on the background spacetime, we can regard the matter on the 4-brane as a small perturbation to the spacetime. Under this assumption, the background solutions are static, satisfying the Einstein's equations without the energy-momentum tensor of the matter. In this case, the bulk metric will become time-dependent only if we introduce (non)linear perturbations sourced from either the brane's motion or the evolution of the matter on the 4-brane.

Here, we derive one of the static vacuum solutions to the field equations~\eqref{fe1} with $\Lambda<0$:
\begin{equation}\label{bm2}
	\text{d}s^{2}_{D}=A(R)\text{d}R^{2}_{~}+B(R)\text{d}\Omega^{2}_{d-1}-C(R)\text{d}T^{2}_{~}+C(R)\text{d}\Sigma^{2}_{3}\,,
\end{equation}
where
\begin{equation}
	A(R)=\frac{(d+3)(d+2)}{-2\Lambda}\frac{1}{R^{2}_{~}},~~~~~B(R)=\frac{(d+3)(d+2)}{-2\Lambda}R^{2}_{~},~~~~~C(R)=R^{2}_{~}.
\end{equation}
This metric describes an $\text{AdS}_{D}^{~}$ bulk spacetime with a $(d-1)$-dimensional sphere,
\begin{equation}
	\text{d}\Omega^{2}_{d-1}=\text{d}\theta^{2}_{1}+\text{sin}^{2}_{~}\theta_{1}^{~}\text{d}\theta_{2}^{~}+\ldots+\text{sin}^{2}_{~}\theta_{1}^{~}\ldots\text{sin}^{2}_{~}\theta_{d-2}^{~}\text{d}\theta_{d-1}^{~},
\end{equation}
and a 3-dimensional flat subspace,
\begin{equation}
	\text{d}\Sigma^{2}_{3}=\text{d}r^{2}_{~}+r^{2}_{~}(\text{d}\psi^{2}_{~}+\text{sin}^{2}_{~}\psi\,\text{d}\phi^{2}_{~}).
\end{equation}
The static vacuum solution~\eqref{bm2} is smooth in the whole bulk. It does not contribute any singular terms to the field equations. So, for an $\text{AdS}_{D}^{~}$ bulk with a 4-brane, it could be applicable everywhere except on the brane's location $R=R_{0}^{~}$ and $y^{m}_{~}=y^{m}_{0}$. To ensure the self-consistency of the field equations~\eqref{fe1}, we can additionally introduce a nonsmooth part, $\Delta g_{MN}^{~}$, into the metric~\eqref{bm2}. Since the 4-brane is static, these corrections are solely sourced from the confined matter on the 4-brane. They can be regarded as perturbations to the background spacetime and do not govern the spacetime structure at the leading order. In the following, we will reserve the spherical symmetry of the extra space under the perturbations. The corrections are then independent of $y^{m}_{~}$ and could be simplified to
\begin{equation}\label{mc1}
	\Delta g_{mn}^{~}=\mathcal{B}(R,T)\tilde{g}_{mn}^{~},~~~~~\Delta g_{TT}^{~}=\mathcal{C}_{T}^{~}(R,T)\hat{g}_{TT}^{~},~~~~~\Delta g_{ij}^{~}=\mathcal{C}_{s}^{(ij)}(R,T)\hat{g}_{ij}^{~}.
\end{equation}
Here, we assume that the contributions of the matter to the $TT$ and $ij$ components of the background metric~\eqref{bm2} are different, i.e., $\mathcal{C}_{T}^{~}\neq\mathcal{C}_{s}^{(ij)}$. We also set the same correction ($\mathcal{C}_{s}^{(ij)}=\mathcal{C}_{s}^{~}$) to each of the $ij$ components on account of the embedding of a homogeneous and isotropic 4-brane. Note that, the functions $\mathcal{B}$, $\mathcal{C}_{T}^{~}$, and $\mathcal{C}_{s}^{~}$ need not be time-independent because they are sourced by $\rho$ and $p$. In addition, there is no correction to $A$ because $A''$ does not appear in the field equations~\eqref{fe1}. With these metric corrections~\eqref{mc1}, the perturbed field equations reduce to
\begin{subequations}\label{fe2}
	\begin{eqnarray}
		\frac{1}{VM^{D-2}_{~}}\frac{1}{\sqrt{A}}\tilde{T}_{ij}^{(5)}
		 &\!=\!& 
		\bigg[-\frac{\ddot{C}_{s}^{~}}{C_{T}^{~}}+\frac{C_{s}''}{A}+\frac{1}{2}\frac{C_{s}^{~}C_{T}''}{AC_{T}^{~}}-\frac{1}{2}\frac{A'C_{s}'}{A^{2}_{~}}-\frac{1}{4}\frac{C_{s}^{~}A'C_{T}'}{A^{2}_{~}C_{T}^{~}}+\frac{1}{4}\frac{\dot{C}_{s}^{2}}{C_{s}^{~}C_{T}^{~}}+\frac{1}{2}\frac{\dot{C}_{s}^{~}\dot{C}_{T}^{~}}{C_{T}^{2}}\nonumber\\
		 &\!~\!& 
		 -\frac{d-1}{4}\frac{A'\mathfrak{B}'C_{s}^{~}}{A^{2}_{~}\mathfrak{B}}-\frac{1}{4}\frac{C_{s}'^{2}}{AC_{s}^{~}}-\frac{d-1}{2}\frac{C_{s}^{~}\ddot{\mathfrak{B}}}{\mathfrak{B}C_{T}^{~}}+\frac{d-1}{2}\frac{C_{s}^{~}\mathfrak{B}''}{A\mathfrak{B}}-\frac{d-1}{2}\frac{\mathfrak{B}'C_{s}'}{A\mathfrak{B}}\nonumber\\
		 &\!~\!& 
		 +\frac{(d-4)(d-1)}{8}\frac{C_{s}^{~}\mathfrak{B}'^{2}_{~}}{A\mathfrak{B}^{2}_{~}}-\frac{d-1}{2}\frac{\dot{\mathfrak{B}}\dot{C}_{s}^{~}}{\mathfrak{B}C_{T}^{~}}+\frac{d-1}{4}\frac{\dot{\mathfrak{B}}C_{s}^{~}\dot{C}_{T}^{~}}{\mathfrak{B}C_{T}^{2}}+\frac{1}{2}\frac{C_{s}'C_{T}'}{AC_{T}^{~}}\nonumber\\
		 &\!~\!& 
		 +\frac{d-1}{4}\frac{C_{s}^{~}\mathfrak{B}'C_{T}'}{A\mathfrak{B}C_{T}^{~}}-\frac{(d-4)(d-1)}{8}\frac{C_{s}^{~}\dot{\mathfrak{B}}^{2}_{~}}{\mathfrak{B}^{2}_{~}C_{T}^{~}}-\frac{1}{4}\frac{C_{s}^{~}C_{T}'^{2}}{AC_{T}^{2}}+C_{s}^{~}\Lambda
		\bigg]\hat{g}_{ij}^{~}\,,\label{fe2a}\\
		\frac{1}{VM^{D-2}_{~}}\frac{1}{\sqrt{A}}\tilde{T}_{TT}^{(5)}
		 &\!=\!& 
		-\frac{3}{2}\frac{C_{T}^{~}C_{s}''}{AC_{s}^{~}}+\frac{3}{4}\frac{\dot{C}_{s}^{2}}{C_{s}^{2}}+\frac{3}{4}\frac{A'C_{T}^{~}C_{s}'}{A^{2}_{~}C_{s}^{~}}+\frac{d-1}{4}\frac{A'\mathfrak{B}'C_{T}^{~}}{A^{2}_{~}\mathfrak{B}}-\frac{d-1}{2}\frac{\mathfrak{B}''C_{T}^{~}}{A\mathfrak{B}}\nonumber\\
		 &\!~\!& 
		 -\frac{(d-4)(d-1)}{8}\frac{\mathfrak{B}'^{2}_{~}C_{T}^{~}}{A\mathfrak{B}^{2}_{~}}+\frac{(d-2)(d-1)}{8}\frac{\dot{\mathfrak{B}}^{2}_{~}}{\mathfrak{B}^{2}_{~}}+\frac{3(d-1)}{4}\frac{\dot{\mathfrak{B}}\dot{C}_{s}^{~}}{\mathfrak{B}C_{s}^{~}}\nonumber\\
		 &\!~\!& 
		 -\frac{3(d-1)}{4}\frac{\mathfrak{B}'C_{s}'C_{T}^{~}}{A\mathfrak{B}C_{s}^{~}}-C_{T}^{~}\Lambda,\label{fe2b}\\
		-A\Lambda
		 &\!=\!& 
		-\frac{3}{2}\frac{A\ddot{C}_{s}^{~}}{C_{s}^{~}C_{T}^{~}}+\frac{3}{4}\frac{C_{s}'^{2}}{C_{s}^{2}}+\frac{3}{4}\frac{C_{s}'C_{T}'}{C_{s}^{~}C_{T}^{~}}+\frac{3}{4}\frac{A\dot{C}_{s}^{~}\dot{C}_{T}^{~}}{C_{s}^{~}C_{T}^{2}}-\frac{d-1}{2}\frac{A\ddot{\mathfrak{B}}}{\mathfrak{B}C_{T}^{~}}+\frac{d-1}{4}\frac{A\dot{\mathfrak{B}}\dot{C}_{T}^{~}}{\mathfrak{B}C_{T}^{2}}\nonumber\\
		 &\!~\!& 
		 +\frac{(d-2)(d-1)}{8}\frac{A\dot{\mathfrak{B}}^{2}_{~}}{\mathfrak{B}^{2}_{~}C_{T}^{~}}-\frac{3(d-1)}{4}\frac{A\dot{\mathfrak{B}}\dot{C}_{s}^{~}}{\mathfrak{B}C_{s}^{~}C_{T}^{~}}+\frac{3(d-1)}{4}\frac{\mathfrak{B}'C_{s}'}{\mathfrak{B}C_{s}^{~}}\nonumber\\
		 &\!~\!& 
		 +\frac{d-1}{4}\frac{\mathfrak{B}'C_{T}'}{\mathfrak{B}C_{T}^{~}},\label{fe2c}\\
		-\mathfrak{B}\Lambda
		 &\!=\!& 
		-\frac{3}{2}\frac{\mathfrak{B}\ddot{C}_{s}^{~}}{C_{s}^{~}C_{T}^{~}}+\frac{3}{2}\frac{\mathfrak{B}C_{s}''}{AC_{s}^{~}}+\frac{1}{2}\frac{\mathfrak{B}C_{T}''}{AC_{T}^{~}}+\frac{3}{4}\frac{\mathfrak{B}C_{s}'C_{T}'}{AC_{s}^{~}C_{T}^{~}}-\frac{1}{4}\frac{\mathfrak{B}C_{T}'^{2}}{AC_{T}^{2}}+\frac{3}{4}\frac{\mathfrak{B}\dot{C}_{s}^{~}\dot{C}_{T}^{~}}{C_{s}^{~}C_{T}^{2}}\nonumber\\
		 &\!~\!& 
		 -\frac{3}{4}\frac{A'\mathfrak{B}C_{s}'}{A^{2}_{~}C_{s}^{~}}-\frac{1}{4}\frac{A'\mathfrak{B}C_{T}'}{A^{2}_{~}C_{T}^{~}}-\frac{d-2}{4}\frac{A'\mathfrak{B}'}{A^{2}_{~}}-\frac{d-2}{2}\frac{\ddot{\mathfrak{B}}}{C_{T}^{~}}+\frac{d-2}{2}\frac{\mathfrak{B}''}{A}\nonumber\\
		 &\!~\!& 
		 +\frac{(d-5)(d-2)}{8}\frac{\mathfrak{B}_{~}'^{2}}{A\mathfrak{B}}-\frac{3(d-2)}{4}\frac{\dot{\mathfrak{B}}\dot{C}_{s}^{~}}{C_{s}^{~}C_{T}^{~}}+\frac{d-2}{4}\frac{\dot{\mathfrak{B}}\dot{C}_{T}^{~}}{C_{T}^{2}}+\frac{3(d-2)}{4}\frac{\mathfrak{B}'C_{s}'}{AC_{s}^{~}}\nonumber\\
		 &\!~\!&
		 +\frac{d-2}{4}\frac{\mathfrak{B}'C_{T}'}{AC_{T}^{~}},\label{fe2d}\\
		\frac{3}{4}\frac{\dot{C}_{s}^{~}C_{s}'}{C_{s}^{2}}
		 &\!=\!& 
		\frac{3}{2}\frac{\dot{C}_{s}'}{C_{s}^{~}}+\frac{d-1}{2}\frac{\dot{\mathfrak{B}}'}{\mathfrak{B}}-\frac{d-1}{4}\frac{\dot{\mathfrak{B}}\mathfrak{B}'}{\mathfrak{B}^{2}_{~}}-\frac{d-1}{4}\frac{\dot{\mathfrak{B}}C_{T}'}{\mathfrak{B}C_{T}^{~}}-\frac{3}{4}\frac{\dot{C}_{s}^{~}C_{T}'}{C_{s}^{~}C_{T}^{~}},\label{fe2e}
	\end{eqnarray}
\end{subequations}
where we have defined $\mathfrak{B}\equiv B+\mathcal{B}$, $C_{s}^{~}\equiv C+\mathcal{C}_{s}^{~}$, and $C_{T}^{~}\equiv C+\mathcal{C}_{T}^{~}$. Note that in the above field equations, we have integrated over the extra $(d-1)$-dimensional space to eliminate $\delta^{(d-1)}_{~}(y^{m}_{~}-y^{m}_{0})$ in the energy-momentum tensor. $\tilde{T}_{MN}^{(5)}$ is thus an effective energy-momentum tensor defined on the rest 5-dimensional submanifold. The parameter $V$ indicates the volume of the extra $(d-1)$-dimensional space, which is related to the quantities for the extra $(d-1)$-dimensional space, such as the spatial curvature and the matter distribution. Our method of integrating the field equations is equivalent to using the dimensionality reduction directly to the $D$-dimensional action to arrive at an effective 5-dimensional theory. The difference is that in our method, we have an extra field equation to constrain the extra freedom arising from the dimensionality reduction.

\subsection{Parameterization of perturbed solutions}

In the perturbed equations~\eqref{fe2}, the bulk metric~\eqref{bm2} is coupled with its perturbations. Although we could eliminate the background field equations by assuming $\mathcal{B}\ll B$, $\mathcal{C}_{s}^{~}\ll C$, and $\mathcal{C}_{T}^{~}\ll C$, the remaining equations governing the perturbations are still nonlinear. It is not practical to solve them directly. Thus, we would like to first derive the linear perturbations. To linearize the perturbed field equations~\eqref{fe2}, the following decompositions of $\mathcal{B}$, $\mathcal{C}_{s}^{~}$, and $\mathcal{C}_{T}^{~}$ are useful:
\begin{subequations}
	\begin{eqnarray}
		\mathcal{B}
		 &=&  
		\mathcal{B}^{(1)}_{~}+\mathcal{B}^{(2)}_{~}+\mathcal{B}^{(3)}_{~}+\ldots\,,\\
		\mathcal{C}_{s}^{~}
		 &=& 
		\mathcal{C}^{(1)}_{s}+\mathcal{C}^{(2)}_{s}+\mathcal{C}^{(3)}_{s}+\ldots\,,\\
		\mathcal{C}_{T}^{~}
		 &=& 
		\mathcal{C}^{(1)}_{T}+\mathcal{C}^{(2)}_{T}+\mathcal{C}^{(3)}_{T}+\ldots\,,
	\end{eqnarray}
\end{subequations}
where $\mathcal{B}^{(1)}_{~}$, $\mathcal{C}_{s}^{(1)}$, and $\mathcal{C}_{T}^{(1)}$ correspond to the linear perturbations, and the other terms denote the higher-order perturbations. The constraints on $\mathcal{B}^{(1)}_{~}$, $\mathcal{C}_{s}^{(1)}$, and $\mathcal{C}_{T}^{(1)}$ are given by the singular part of the linearized perturbation equations from Eqs.~\eqref{fe2a},~\eqref{fe2b}, and~\eqref{fe2d},
\begin{subequations}
	\begin{eqnarray}
		\frac{d-1}{2}\frac{(\mathcal{B}^{(1)}_{~})''}{B}
		  &=& 
		-\frac{3}{2}\frac{(\mathcal{C}_{s}^{(1)})''}{C}
		-\frac{\sqrt{A}\,\rho\,\delta(R-R_{0}^{~})}{VM^{D-2}_{~}},\\
		\frac{d-1}{2}\frac{(\mathcal{B}^{(1)}_{~})''}{B}
		 &=& 
		-\frac{(\mathcal{C}_{s}^{(1)})''}{C}-\frac{1}{2}\frac{(\mathcal{C}_{T}^{(1)})''}{C}-\frac{\sqrt{A}\,p\,\delta(R-R_{0}^{~})}{VM^{D-2}_{~}},
		\\
		\frac{d-2}{2}\frac{(\mathcal{B}^{(1)}_{~})''}{B}
		 &=& 
		-\frac{3}{2}\frac{(\mathcal{C}_{s}^{(1)})''}{C}-\frac{1}{2}\frac{(\mathcal{C}_{T}^{(1)})''}{C}.
	\end{eqnarray}
\end{subequations}
So the local solutions of $\mathcal{B}^{(1)}_{~}$, $\mathcal{C}_{s}^{(1)}$, and $\mathcal{C}_{T}^{(1)}$ near the 4-brane can be approximated as
\begin{subequations}\label{lp1}
	\begin{eqnarray}
		\mathcal{B}^{(1)}_{~}
		 &\!=\!& 
		-\frac{\rho-3p}{VM^{D-2}_{~}}\frac{2}{d+2}\bigg[\frac{(d+3)(d+2)}{-2\Lambda}\bigg]^{\frac{3}{2}}_{~}R\,|R-R_{0}^{~}|,
		\\
		\mathcal{C}_{T}^{(1)}
		 &\!=\!& 
		\frac{(d+1)\rho+3p}{VM^{D-2}_{~}}\frac{2}{d+2}\sqrt{\frac{(d+3)(d+2)}{-2\Lambda}}R\,|R-R_{0}^{~}|,
		\\
		\mathcal{C}_{s}^{(1)}
		 &\!=\!& 
		-\frac{\rho+(d-1)p}{VM^{D-2}_{~}}\frac{2}{d+2}\sqrt{\frac{(d+3)(d+2)}{-2\Lambda}}R\,|R-R_{0}^{~}|.
	\end{eqnarray}
\end{subequations}
Obviously, they are the linear-in-$|R-R_{0}^{~}|$ corrections to the metric. One could check that it is hard to describe the normal expansion of the 4-brane solely using the linear perturbation equations in this model. So it is necessary to introduce corrections beyond the order of $\mathcal{O}(|R-R_{0}^{~}|^{2}_{~})$ to the field equations.
These corrections include the higher-order metric perturbations (e.g. $\mathcal{B}^{(2)}_{~}$ and $\mathcal{B}^{(3)}_{~}$) sourced by lower-order perturbations and the higher-order products (e.g. $\mathcal{B}^{(1)}_{~}\mathcal{B}^{(1)}_{~}$ and $\mathcal{B}^{(1)}_{~}\mathcal{B}^{(2)}_{~}$) directly constructed by the lower ones. They do not contribute singular terms near the brane, but smooth nonlinear terms to the field equations. In the next section, we will show how these nonlinear perturbations stabilize the model and help us to recover the standard cosmology on the 4-brane.

Taking all these higher-order corrections into account, we finally arrive at the following parameterization of the perturbed metric:
\begin{equation}\label{bm3}
	\text{d}s_{D}^{2}
	 =
	 A\,\text{d}R^{2}_{~}+B\,e^{\mathcal{B}_{0}^{~}|R-R_{0}^{~}|}_{~}\,\text{d}\Omega_{d-1}^{2}-C\,e^{\mathcal{C}_{T0}^{~}|R-R_{0}^{~}|}_{~}\,\text{d}T^{2}_{~}+\,C\,e^{\mathcal{C}_{s0}^{~}|R-R_{0}^{~}|}_{~}\,\text{d}\Sigma^{2}_{3}\,,
\end{equation}
where
\begin{subequations}
	\begin{eqnarray}
		\mathcal{B}_{0}^{~}
		 &=& 
		-\frac{\rho-3p}{VM^{D-2}_{~}}\frac{2}{d+2}\sqrt{\frac{(d+3)(d+2)}{-2\Lambda}}\frac{1}{R}\,,\\
		\mathcal{C}_{T0}^{~}
		 &=& 
		\frac{(d+1)\rho+3p}{VM^{D-2}_{~}}\frac{2}{d+2}\sqrt{\frac{(d+3)(d+2)}{-2\Lambda}}\frac{1}{R}\,,\\
		\mathcal{C}_{s0}^{~}
		 &=& 
		-\frac{\rho+(d-1)p}{VM^{D-2}_{~}}\frac{2}{d+2}\sqrt{\frac{(d+3)(d+2)}{-2\Lambda}}\frac{1}{R}\,.~
	\end{eqnarray}
\end{subequations}
Note that since each exponent in the metric components yields a constant factor at infinity, which would prevent~\eqref{bm3} from recovering the background metric~\eqref{bm2}, one can introduce the coordinate transformations to solve the problem. Such coordinate transformations only need to eliminate those constant factors, so they are trivial and easily to be found. Under these transformations, the nonsmooth part of the parameterization still satisfies the boundary condition of the field equations at $R=R_{0}^{~}$, and gives a non-trivial correction to the bulk metric near the brane. Far away from the brane, the non-trivial correction is carried by the higher-order terms of the parameterization,
while the background bulk metric remains the leading order. In fact, the parameterization~\eqref{bm3} extents the validity of the perturbed metric solution into the whole bulk, since it carries all the nonlinear-in-$|R-R_{0}^{~}|$ terms. Although a more accurate matching prefers a more general parameterization, it will make the calculation much more complicated. Next, our discussions will be based on the parameterization~\eqref{bm3}.

\section{Dynamical 4-brane}\label{sec3}

In the last section, we assume that the background spacetime keeps static under the embedding of the 4-brane. It is a very strong restriction, and could make it hard to describe a normal expansion of the 4-brane. Indeed, the matter on the 4-brane could result in the motion of the 4-brane in the bulk, and could cause the spacetime time-dependent at the $\sim |R-R_{0}^{~}|^{0}_{~}$ order as well. In this section, we will show how the motion of a 4-brane is driven by the energy density and pressure of the matter confined on it, and how the balancing between the bulk and bare cosmological constants recovers the standard cosmology.

\subsection{Stable time-dependent generalization}

We let the 4-brane move in $R$ direction while still fixing it in $y^{m}_{~}$ direction for convenience. The location of the 4-brane is then described by
\begin{equation}\label{bp2}
	R=\mathcal{R}(T),~~~y^{m}_{~}=y^{m}_{0}\,.
\end{equation}
With this condition, we can construct a series of spacelike normal vectors $n_{M}^{(\mathfrak{m})}$'s to each of the $\mathfrak{m}$-dimensional hypersurface by
\begin{equation}
	n_{M}^{(\mathfrak{m})}\equiv\nabla_{M}^{~}f^{\mathfrak{m}}_{~},
\end{equation}
where $\mathfrak{m}$ ranges from $4$ to $D-1$. According to the condition~\eqref{bp2}, the parameter function for the 4-brane can be given by
\begin{equation}
	f^{\mathfrak{m}}_{~}=y^{\mathfrak{n}}_{0}\,\,\,\text{for}\,\,\, \mathfrak{m}>4
\end{equation}
and
\begin{equation}
	f^{\mathfrak{m}}_{~}=R-\mathcal{R}(T)\,\,\,\text{for}\,\,\, \mathfrak{m}=4.
\end{equation}
Here, we have defined $\mathfrak{n}=\mathfrak{m}-3$. Then, the normalization $\tilde{n}_{M}^{(\mathfrak{m})}\tilde{n}^{(\mathfrak{m}),M}_{\color{white}{none}}=1$ yields
\begin{equation}
	\tilde{n}_{M}^{(\mathfrak{m})}=\delta_{~M}^{\mathfrak{n}}\sqrt{g_{\mathfrak{n}\mathfrak{n}}^{~}}\,\,\,\text{for}\,\,\, \mathfrak{m}>4
\end{equation}
and
\begin{equation}
	\tilde{n}_{M}^{(4)}=\bigg(-\sqrt{\frac{AC}{C-A\dot{\mathcal{R}}^{2}_{~}}},0,0,0,\sqrt{\frac{AC}{C-A\dot{\mathcal{R}}^{2}_{~}}},0,\ldots,0\bigg),
\end{equation}
where dots denote derivatives with respect to the bulk time $T$. With these unit normal vectors $\tilde{n}_{M}^{(\mathfrak{m})}$, the projection tensor for the 4-brane turns into
\begin{equation}
	\tilde{h}_{MN}^{~}
	 =g_{MN}^{~}-\sum_{\mathfrak{m}=4}^{D-1}\tilde{n}_{M}^{(\mathfrak{m})}\tilde{n}_{N}^{(\mathfrak{m})}
	 =g_{MN}^{~}-\tilde{n}_{M}^{(4)}\tilde{n}_{N}^{(4)}-\delta_{~M}^{m}\delta_{~M}^{n}\,g_{mn}^{~}\,.
\end{equation}
For the observer on the 4-brane, the unit velocity vector can be denoted as
\begin{equation}
	\tilde{u}^{M}_{~}=\bigg(\frac{1}{\sqrt{C-A\dot{\mathcal{R}}^{2}_{~}}},0,0,0,\frac{\dot{\mathcal{R}}}{\sqrt{C-A\dot{\mathcal{R}}^{2}_{~}}},0,\ldots,0\bigg),
\end{equation}
which involves the motion of the 4-brane in $R$ direction. One can check that the velocity vector is already projected on the 4-brane and orthogonal to the brane's normal vector by $\tilde{h}_{MN}^{~}\tilde{u}^{M}_{~}=0$ and $\tilde{n}_{M}^{(\mathfrak{m})}\tilde{u}^{M}_{~}=0$. Recalling the condition~\eqref{bp2}, we can construct the energy-momentum tensor of the matter on the moving 4-brane as follows:
\begin{equation}
	T_{MN}^{~}=\big[(\rho+p)\tilde{u}_{M}^{~}\tilde{u}_{N}^{~}+p\,\tilde{h}_{MN}^{~}\big]\delta(R-\mathcal{R})\delta^{(d-1)}_{~}(y^{m}_{~}-y^{m}_{0}).
\end{equation}
Plugging the condition~\eqref{bp2} into the line element~\eqref{bm2}, the induced metric $\tilde{g}_{\mu\nu}^{(4)}$ that the 4-dimensional fields couple to is also modified to
\begin{equation}\label{im1}
	\text{d}s_{4}^{2}=-\mathcal{R}^{2}_{~}\mathcal{H}^{2}_{~}\text{d}T^{2}_{~}+\mathcal{R}^{2}_{~}\text{d}\Sigma_{3}^{2}
\end{equation}
with
\begin{equation}\label{mH1}
	\mathcal{H}^{2}_{~}=1-\frac{(d+3)(d+2)}{-2\Lambda}\frac{\dot{\mathcal{R}}^{2}_{~}}{\mathcal{R}^{4}_{~}}.
\end{equation}
In this case, a time-dependent generalization of the perturbed metric~\eqref{bm3} is written as
\begin{equation}\label{bm4}
	\text{d}s_{D}^{2}=\tilde{\mathfrak{A}}\text{d}R^{2}_{~}+\tilde{\mathfrak{B}}\text{d}\Omega_{d-1}^{2}-\tilde{C}_{T}^{~}\text{d}T^{2}_{~}+\tilde{C}_{s}^{~}\text{d}\Sigma_{3}^{2},
\end{equation}
where
\begin{subequations}
	\begin{eqnarray}
		\tilde{\mathfrak{B}}
		 &=& 
		 B\,\mathcal{B}_{1}^{~}(R,T)\,e^{\mathcal{B}_{0}^{~}|R-\mathcal{R}|}_{~},\\
		\tilde{C}_{T}^{~}
		 &=& 
		 C\,\mathcal{C}_{T1}^{~}(R,T)\,e^{\mathcal{C}_{T0}^{~}|R-\mathcal{R}|}_{~},\\
		\tilde{C}_{s}^{~}
		 &=& 
		 C\,\mathcal{C}_{s1}^{~}(R,T)\,e^{\mathcal{C}_{s0}^{~}|R-\mathcal{R}|}_{~}.
	\end{eqnarray}
\end{subequations}
Note that on account of the appearance $\ddot{A}$ in the $ij$ and $mn$ components of the field equations, the $RR$ component of the metric has been generalized to
\begin{equation}
	\tilde{\mathfrak{A}}=A\,\mathcal{A}_{1}^{~}(R,T)\,e^{\mathcal{A}_{0}^{~}|R-\mathcal{R}|}_{~},
\end{equation}
where the smooth function $\mathcal{A}_{0}^{~}$ is sourced from the brane matter. Unlike $\mathcal{B}_{0}^{~}$, $\mathcal{C}_{T0}^{~}$, and $\mathcal{C}_{s0}^{~}$, $\mathcal{A}_{0}^{~}$ cannot be constrained by the joint condition for the case of a static 4-brane due to the absence of $A''$. In other words, $\mathcal{A}_{0}^{~}$ should vanish when the 4-brane is fixed in the bulk. A tricky way is to parameterize it as $\mathcal{A}_{0}^{~}=\mathcal{A}_{0}^{(1)}(\rho,p)\dot{\mathcal{R}}+\mathcal{A}_{0}^{(2)}(\rho,p)\dot{\mathcal{R}}^{2}_{~}+\ldots\,$, where the coefficients should satisfy the field equations. Compared with the metric~\eqref{bm3} in the last section, the bulk spacetime now becomes time-dependent at the $|R-\mathcal{R}|^{0}_{~}$ order with the new functions $\mathcal{A}_{1}^{~}$, $\mathcal{B}_{1}^{~}$, $\mathcal{C}_{T1}^{~}$, and $\mathcal{C}_{s1}^{~}$. These functions are smooth and dominated by $\rho$, $p$, and the motion of the 4-brane. As before, we can still use the singular terms of the field equations to derive expressions for them. What is different is that field equations now include extra singular terms arising from, for example, second time derivatives acting on $\tilde{\mathfrak{A}}$, $\tilde{\mathfrak{B}}$, and other variables.

Taking the generalized metric~\eqref{bm4} into account, the singular parts of linearized perturbation equations at the $|R-\mathcal{R}|^{1}_{~}$ order become
\begin{subequations}\label{Ijc2}
	\begin{eqnarray}
		\frac{p}{VM^{D-2}_{~}}\frac{\mathcal{R}^{4}_{~}}{C^{2}_{~}}\sqrt{A}\,\mathcal{H}\delta(R-\mathcal{R})
		 &=& 
		\Big(\frac{d-1}{2}\mathcal{B}_{0}^{~}\mathcal{B}_{1}^{~}+\mathcal{C}_{s0}^{~}\mathcal{C}_{s1}^{~}\Big)\Big(1-\frac{A}{C}\dot{\mathcal{R}}^{2}_{~}\Big)\delta(R-\mathcal{R})\nonumber\\
		 &~& 
		 +\frac{1}{2}\mathcal{C}_{T0}^{~}\mathcal{C}_{T1}^{~}\delta(R-\mathcal{R})-\frac{1}{2}\frac{A}{C}\mathcal{A}_{0}^{~}\mathcal{A}_{1}^{~}\dot{\mathcal{R}}^{2}_{~}\delta(R-\mathcal{R}),\\
		-\frac{\rho}{VM^{D-2}_{~}}\frac{\mathcal{R}^{4}_{~}}{C^{2}_{~}}\sqrt{A}\,\mathcal{H}\delta(R-\mathcal{R})
		 &=& 
		\Big(\frac{d-1}{2}\mathcal{B}_{0}^{~}\mathcal{B}_{1}^{~}+\frac{3}{2}\mathcal{C}_{s0}^{~}\mathcal{C}_{s1}^{~}\Big)\Big(1-\frac{A}{C}\dot{\mathcal{R}}^{2}_{~}\Big)\delta(R-\mathcal{R}),\\
		\frac{1}{2}\frac{A}{C}\mathcal{A}_{0}^{~}\mathcal{A}_{1}^{~}\dot{\mathcal{R}}^{2}_{~}\delta(R-\mathcal{R})
		 &=& 
		\Big(\frac{d-2}{2}\mathcal{B}_{0}^{~}\mathcal{B}_{1}^{~}+\frac{3}{2}\mathcal{C}_{s0}^{~}\mathcal{C}_{s1}^{~}\Big)\Big(1-\frac{A}{C}\dot{\mathcal{R}}^{2}_{~}\Big)\delta(R-\mathcal{R})\nonumber\\
		 &~& 
		 +\frac{1}{2}\mathcal{C}_{T0}^{~}\mathcal{C}_{T1}^{~}\delta(R-\mathcal{R}).
	\end{eqnarray}
\end{subequations}
Here, the number of the independent equations are less than that of the functions we should constrain on the boundary. We then set $\mathcal{A}_{1}^{~}=1$, and let $\mathcal{C}_{T1}^{~}=(1-A\dot{\mathcal{R}}^{2}_{~}/C)^{1/2}_{~}+(A\dot{\mathcal{R}}^{2}_{~}/C)^{2}_{~}$ by choosing the gauge for the bulk time. Since we assume that the 4-brane has a little back-reaction to the bulk spacetime, the functions $\mathcal{B}_{1}^{~}$, $\mathcal{C}_{s1}^{~}$, and $\mathcal{C}_{T1}^{~}$ should follow $\mathcal{B}_{1}^{~}=\mathcal{C}_{s1}^{~}=\mathcal{C}_{T1}^{~}=1$ far away from the 4-brane. The local solutions near the 4-brane, consistent with our assumption and the boundary conditions~\eqref{Ijc2}, are given by
\begin{equation}
	\mathcal{B}_{1}^{~}
		 =
		\mathcal{C}_{s1}^{~}=\Big(1-\frac{A}{C}\dot{\mathcal{R}}^{2}_{~}\Big)^{-1/2}_{~},~~~~~
		\mathcal{A}_{0}^{~}
		 =
		\mathcal{A}_{0}^{(2)}\dot{\mathcal{R}}^{2}_{~}=\frac{A}{C}\mathcal{C}_{T0}^{~}\dot{\mathcal{R}}^{2}_{~}.
\end{equation}
So, recalling Eq.~\eqref{mH1}, we have $\mathcal{B}_{1}^{~}=\mathcal{C}_{s1}^{~}=\mathcal{C}_{T1}^{-1}=1/\mathcal{H}$, $\mathcal{A}_{0}^{~}=\mathcal{C}_{T0}^{~}(1-\mathcal{H}^{2}_{~})$, and $\mathcal{C}_{T1}^{~}=\mathcal{H}+(1-\mathcal{H}^{2}_{~})^{2}_{~}$ on the 4-brane. One can check that $\mathcal{B}_{1}^{~}=\mathcal{C}_{s1}^{~}=\mathcal{C}_{T1}^{~}=1$ when the 4-brane is fixed in the bulk. That is to say, when the contributions from $\rho$ and $p$ to the brane's motion is negligible, the background spacetime can be considered static in this model. Once again, the contributions from the brane's back-reaction are treated as perturbations in our model. We prefer the scenario in which the model can maintain stability under these perturbations. It can be proved that the stability of the model requires $A\dot{\mathcal{R}}^{2}_{~}/C\ll1$ near the 4-brane. Reminding that our induced metric~\eqref{im1} on the 4-brane is consistent with the Friedmann-Lema\^{\i}tre-Robertson-Walker (FLRW) metric with vanishing curvature:
\begin{equation}\label{FLRW1}
	\text{d}s_{4}^{2}=-\text{d}t^{2}_{~}+a(t)^{2}_{~}\text{d}\Sigma_{3}^{2},
\end{equation}
where $t$ is the cosmic time and $a(t)$ is the scale factor. So the brane's location~\eqref{bp2} in $R$ direction is related to the expansion of the 4-brane itself through
\begin{equation}\label{bcr1}
	a^{2}_{~}=\mathcal{R}^{2}_{~}.
\end{equation}
In this case, comparing the two metrics, one can obtain a mapping between the bulk time and the cosmic time,
\begin{equation}\label{bcr2}
	\text{d}t^{2}_{~}
		=
		\mathcal{R}^{2}_{~}\mathcal{H}^{2}_{~}\text{d}T^{2}_{~}.
\end{equation}
With the relations~\eqref{bcr1} and~\eqref{bcr2}, it is found that, near the brane, $A\dot{\mathcal{R}}^{2}_{~}/C$ satisfies
\begin{equation}
	\frac{A(\mathcal{R})}{C(\mathcal{R})}\dot{\mathcal{R}}^{2}=\frac{(d+2)(d+3)}{-2\Lambda}H^{2}_{~}\Big[1+\frac{(d+2)(d+3)}{-2\Lambda}H^{2}_{~}\Big]^{-1}_{~},
\end{equation}
where $H$ is the Hubble parameter. Thus, in the low-energy case $\left(\frac{(d+3)(d+2)}{-2\Lambda}H^{2}_{~} \ll 1\right)$, the perturbed metric components near the 4-brane become 
\begin{subequations}
\begin{eqnarray}
    \tilde{\mathfrak{A}}
	&\sim& A,\\
	\tilde{\mathfrak{B}}
	&\sim& B\Big[1+\frac{(d+2)(d+3)}{-4\Lambda}H^{2}_{~}+\mathcal{O}(H^{4}_{~}/\Lambda^{2}_{~})\Big],\\
	\tilde{\mathcal{C}}_{T}^{~}
	&\sim& C\Big[1-\frac{(d+2)(d+3)}{-4\Lambda}H^{2}_{~}+\mathcal{O}(H^{4}_{~}/\Lambda^{2}_{~})\Big],\\
	\tilde{\mathcal{C}}_{s}^{~}
	&\sim& C\Big[1+\frac{(d+2)(d+3)}{-4\Lambda}H^{2}_{~}+\mathcal{O}(H^{4}_{~}/\Lambda^{2}_{~})\Big].
\end{eqnarray}
\end{subequations}
The corrections to the metric consistently remain small. They could be treated as small perturbations to the background spacetime. These perturbations do not exhibit divergence over time, so the model is always stable, which is consistent with our assumptions. The result is expected, since we have introduced nonlinear perturbations through the parameterization~\eqref{bm3} to cure the instability of the linear perturbation~\eqref{lp1} that possibly occurs under our dynamic generalization to the perturbed metric.

\subsection{Cosmology on the 4-brane}

The dynamics of the 4-brane is described by the non-singular part of the field equations near the 4-brane. Substituting the metric~\eqref{bm4} into the perturbation equations~\eqref{fe2}, we have
\begin{subequations}\label{de1}
	\begin{eqnarray}
		\mathcal{F}_{1}^{~}\frac{\ddot{\mathcal{R}}}{\mathcal{R}}
		 &=& 
		\big[\mathcal{F}_{2}^{~}(1-\mathcal{H}_{~}^{2})-\mathcal{F}_{3}^{~}\mathcal{H}^{2}_{~}-\mathcal{F}_{4}^{~}\mathcal{H}_{~}^{2}(1-\mathcal{H}_{~}^{2})\big]\mathcal{R}_{~}^{2}-\mathcal{F}_{5}^{~}(1-\mathcal{H}_{~}^{2})\mathcal{H}\frac{4\mathcal{H}-1-4\mathcal{H}_{~}^{3}}{\mathcal{H}+(1-\mathcal{H}_{~}^{2})^{2}_{~}}\mathcal{R}_{~}^{2},\nonumber\\
		 &~& 
		 \\
		-\frac{\ddot{\mathcal{R}}}{\mathcal{R}^{3}_{~}}
		 &=& 
		\frac{\mathcal{F}_{8}^{~}+\mathcal{F}_{9}^{~}\mathcal{H}^{2}_{~}+\mathcal{F}_{10}^{~}(1-\mathcal{H}_{~}^{2})}{\mathcal{F}_{6}^{~}-\mathcal{F}_{7}^{~}(1-\mathcal{H}_{~}^{2})}\big[\mathcal{H}+(1-\mathcal{H}_{~}^{2})^{2}_{~}\big]+\mathcal{F}_{11}^{~}\frac{\mathcal{H}^{2}_{~}}{1-\mathcal{H}_{~}^{2}}\frac{\mathcal{H}+(1-\mathcal{H}_{~}^{2})^{2}_{~}}{\mathcal{F}_{6}^{~}-\mathcal{F}_{7}^{~}(1-\mathcal{H}_{~}^{2})}\nonumber\\
		 &~& 
		+\frac{\mathcal{F}_{12}^{~}+\mathcal{F}_{13}^{~}(1-\mathcal{H}_{~}^{2})}{\mathcal{F}_{6}^{~}-\mathcal{F}_{7}^{~}(1-\mathcal{H}_{~}^{2})}\mathcal{H}^{2}_{~},
	\end{eqnarray}
\end{subequations}
where $\mathcal{F}_{i}^{~}$'s are the parameter functions in terms of $d$, $\Lambda$, $\rho$, and $p$. The dynamics of the 4-brane is governed by $\rho$ and $p$. Note that with~\eqref{bcr1} and~\eqref{bcr2}, the variables of the 4-brane could be converted into the cosmological quantities through
\begin{subequations}\label{bbm1}
	\begin{eqnarray}
		\frac{1}{\mathcal{H}^{2}_{~}}
		 &=& 
		1+\frac{(d+3)(d+2)}{-2\Lambda}H^{2}_{~},\\
		\frac{\ddot{\mathcal{R}}}{\mathcal{R}}
		 &=& 
		\bigg[\frac{\partial^{2}_{~}\mathcal{R}}{\partial t^{2}_{~}}\frac{1}{\mathcal{R}}+H^{2}_{~}+2\frac{(d+3)(d+2)}{-2\Lambda}H^{4}_{~}\bigg]\mathcal{R}^{2}_{~}\mathcal{H}^{4}_{~}.~~~~~~~
	\end{eqnarray}
\end{subequations}
So the cosmology on the 4-brane could be derived from the brane's dynamic equations~\eqref{de1}. However, it is not easy to recover the standard form of cosmology directly because of the higher-order terms $\mathcal{O}(\mathcal{H}^{3}_{~})$ in the equations~\eqref{de1}. We therefore employ the following parameterization for each of the cosmological quantities:
\begin{equation}\label{p2}
	\mathcal{Q}=\mathcal{Q}^{(0)}_{~}+\mathcal{Q}^{(1)}_{~}(\rho,p)+\mathcal{Q}^{(2)}_{~}(\rho^{2}_{~},p^{2}_{~})+\ldots\,,
\end{equation}
where we assume that the effects of extra dimensions are entirely incorporated within the higher-order terms $\mathcal{O}(\rho^{2}_{~},p^{2}_{~})$. The lower-order terms should satisfy the present cosmological observations. They should be consistent with the prediction from the standard cosmology on the 4-brane. Thus, they would be the dominant terms in the expression~\eqref{p2} for the late universe. On the contrary, the higher-order terms are expected to contribute considerable corrections primarily during the early universe. They are therefore treated as perturbations to $Q$ in the context of the late universe.

Substituting Eqs.~\eqref{bbm1} and~\eqref{p2} into Eq.~\eqref{de1}, we find the following modified Friedmann equation:
\begin{equation}\label{Fe1}
	\frac{\partial^{2}_{~}\mathcal{R}}{\partial t^{2}_{~}}\frac{1}{\mathcal{R}}+H^{2}_{~}
	=
	\frac{\rho-3p}{6M_{\text{pl}}^{2}}+\frac{\mathcal{P}_{1}^{~}\,p^{2}_{~}+\mathcal{P}_{2}^{~}\,p\,\rho+\mathcal{P}_{3}^{~}\,\rho^{2}_{~}}{M^{2D-4}_{~}V^{2}_{~}}+\mathcal{O}(\rho^{3}_{~},p^{3}_{~}),
\end{equation}
where $\mathcal{P}_{1}^{~}$, $\mathcal{P}_{2}^{~}$, and $\mathcal{P}_{3}^{~}$ are parameter functions only related to the dimensionality of the extra space.
The bulk cosmological constant has been finely tuned to
\begin{equation}\label{ft1}
	\Lambda
	=
	-\frac{(7 d+17) (d+3)}{36 (d+2)}\mathcal{P}_{4}^{~}\lambda= 
	-\frac{(d+3)(d+2)}{2}\mathcal{P}_{4}^{2}\bigg(\frac{M^{D-2}_{~}V}{M_{\text{pl}}^{2}}\bigg)^{2}_{~}
\end{equation}
to reproduce the predictions of general relativity at the leading order (see the first term on the r.h.s. of Eq.~\eqref{Fe1}). Here, $\mathcal{P}_{4}^{~}$ is also the parameter function solely dependent on $d$.
\begin{figure}[!htb]
\center{
\subfigure[]{\includegraphics[width=6cm]{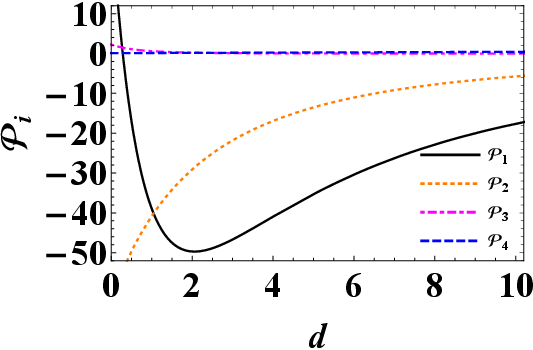}\label{numdp1}}
\quad\quad
\subfigure[]{\includegraphics[width=6cm]{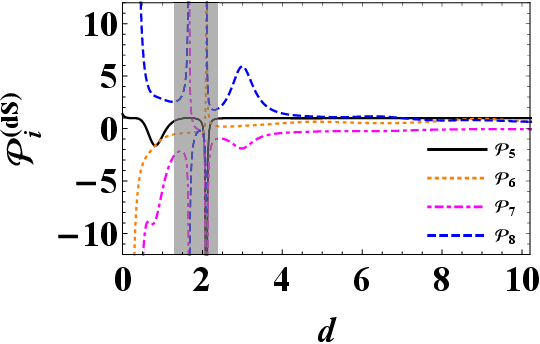}\label{numdp2}}
}
\caption{Parameter functions $\mathcal{P}_{i}^{~}$'s with respect to the dimension of the extra space. (a) Analytical results for the 4-brane with a vanishing effective cosmological constant. (b) Fitting results for the $\text{dS}_{4}^{~}$ brane. The gray area indicates the region where singular behaviors appear in the numerical results.}
\label{numdp}
\end{figure}
We plot Fig.~\ref{numdp1} to show the behaviors of $\mathcal{P}_{i}^{~}$'s. The parameter $\lambda$ is supposed to be the bare cosmological constant (on the 4-brane) introduced by $\rho\rightarrow\rho+M_{\text{pl}}^{2}\lambda$. As is shown in Eq.~\eqref{Fe1}, the modified Friedmann equation adopts a vanishing effective cosmological constant $\Lambda_{\text{eff}}^{~}$ on the 4-brane, which, as we will see later, is related to both the bulk cosmological constant and the bare cosmological constant. It can describe the normal expansion of the late universe with $\Lambda_{\text{eff}}^{~}=0$ through the leading-order term, and provides nontrivial corrections to the expansion of the universe during the early universe.

For the (A)dS 4-brane, it is more difficult to simplify the modified Friedmann equation even if we apply the parameterization~\eqref{p2}. Here, we can introduce the low-energy approximation to Eqs.~\eqref{de1}.
Then, with the following fine-tuning on the bulk cosmological constant and the bare cosmological constant:
\begin{equation}\label{bcbc1}
	\Lambda
		 = 
		-\frac{(d+3)(d+2)}{2}\bigg(\frac{M^{D-2}_{~}V}{M_{\text{pl}}^{2}}\bigg)^{2}_{~}\mathcal{Y}^{2}_{~},~~~~~\lambda
		 = 
		\frac{-2\Lambda}{(d+3)(d+2)}\mathcal{X},
\end{equation}
the modified Friedmann equation becomes
\begin{equation}\label{Fe2}
	\frac{\partial ^2\mathcal{R}}{\partial t^2}\frac{1}{\mathcal{R}}+H^2
	 = 
	\frac{2}{3}\Lambda _{\text{eff}}^{~}+\frac{\rho -3 p}{6 M_{\text{pl}}^2}+\frac{\mathcal{P}_6^{~}\,p^2_{~} +\mathcal{P}_7^{~}\,p \,\rho +\mathcal{P}_8^{~}\, \rho ^2}{\mathcal{P}_5^{~}  M^{2 D-4}_{~}V^2_{~}}+\mathcal{O}(\rho ^3_{~},p^3_{~}),
\end{equation}
where $\mathcal{P}_{5}^{~}$, $\mathcal{P}_{6}^{~}$, $\mathcal{P}_{7}^{~}$, and $\mathcal{P}_{8}^{~}$ are parameter functions in terms of $\mathcal{X}$, $\mathcal{Y}$, and $d$. The parameter functions $\mathcal{X}$ and $\mathcal{Y}$ could be numerically solved from Eq.~\eqref{de1}.
\begin{figure}[!htb]
\center{
\subfigure[]{\includegraphics[width=6cm]{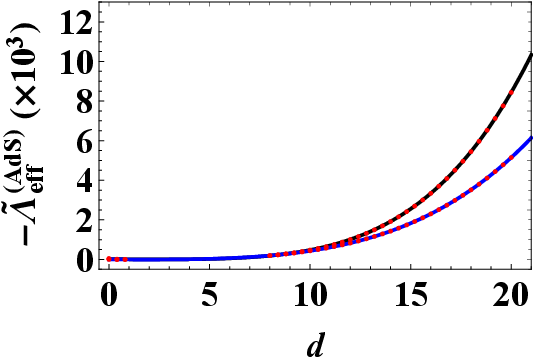}\label{numads1}}
\quad\quad
\subfigure[]{\includegraphics[width=6cm]{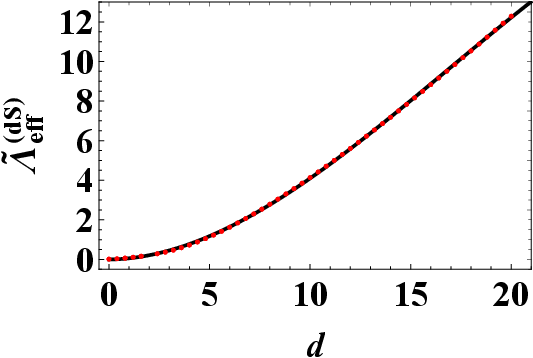}\label{numds1}}
}
\caption{Numerical results of the effective cosmological constant with respect to the dimension of the extra space. (a) For an $\text{AdS}_{4}^{~}$ brane, the blue and black lines correspond to two different branches of fitting results. (b) For a $\text{dS}_{4}^{~}$ brane, the black line is the unique fitting result. In both two cases, red points are the numerical results. The rescaled effective cosmological constant is dimensionless, and is defined by  $\tilde{\Lambda}_{\text{eff}}^{~}=\big(\frac{M_{\text{pl}}^{2}}{M^{D-2}_{~}V}\big)^{2}_{~}\Lambda_{\text{eff}}^{~}$.}
\label{num1}
\end{figure}
The effective cosmological constant is related to $\Lambda$ and $\lambda$ by
\begin{equation}
	\Lambda_{\text{eff}}^{~}=\frac{-6\Lambda}{(d+3)(d+2)}\frac{1+\mathcal{X}\mathcal{Y}}{(d-3)-\mathcal{X}\mathcal{Y}}\,.
\end{equation}
As is shown in Fig.~\ref{numads1}, it has two branches of negative solutions, referring to two kinds of $\text{AdS}_{4}^{~}$ brane in the model. Although our numerical calculation fails to read out the result for $\Lambda_{\text{eff}}^{~}<0$ when $1<d<8$, we can introduce the fitting function matching the numerical result to give a semi-analytical analysis. Our model also allows the embedding of a $\text{dS}_{4}^{~}$ brane in the bulk. As is shown in Fig.~\ref{numds1}, the effective cosmological constant is positive for $\Lambda$ and $\lambda$ given by~\eqref{bcbc1}. Here we only show our fitting results of the parameter functions $\mathcal{P}_{5}^{~}$, $\mathcal{P}_{6}^{~}$, $\mathcal{P}_{7}^{~}$, and $\mathcal{P}_{8}^{~}$ for the $\text{dS}_{4}^{~}$ brane in Fig.~\ref{numdp2}.

\section{Conclusion and discussion}\label{sec4}

In the existing literature of brane cosmology, field equations was solved under the ``bulk-based'' approach, where the bulk spacetime keeps static and smooth under the motion of the 4-brane. Although the standard dS expansion of the brane is recovered by the balancing between the bulk and bare cosmological constants, the brane's back-reaction to the bulk spacetime has been ignored by considering the Israel condition. We should note that, the 4-brane will contribute singular terms to the field equations through its nonvanishing energy-momentum tensor. These singular terms are not consistent with the smooth geometry of the spacetime. It reveals the necessity of introducing nonsmooth corrections to the bulk metric.
In this paper, we employed a new approach to construct a dynamical braneworld model with the 4-brane having nonlinear back-reaction to the $\text{AdS}_{D}^{~}$ spacetime.

We considered a $D$-dimensional spacetime with a bulk cosmological constant. The ordinary matter is confined on a 4-brane embedded in this spacetime. The underlying gravity is described by the $D$-dimensional general relativity. First, we assume that the 4-brane is fixed in the bulk through the condition~\eqref{bp1}. The singularity of the energy-momentum tensor~\eqref{emt1} then makes the metric nonsmooth on the location of the 4-brane. Since the 4-brane is treated as a small perturbation, it only contributes nonsmooth corrections to the metric at the perturbation level. In this case, the background field equations are the vacuum Einstein equations with a smooth background metric. The local solution~\eqref{lp1} of the linearized metric perturbation in the $\text{AdS}_{D}^{~}$ bulk~\eqref{bm2} was derived from the singularity part of the linearized perturbation equations. Analyzing the behavior of the nonlinear perturbations, we found the parameterization~\eqref{bm3} of the nonlinearly perturbed metric. This parameterization extends the validity of the metric perturbation solution to the whole bulk by taking all the nonlinear-in-$|R-R_{0}^{~}|$ terms into account. Under a coordinate transformation, it can recover the background metric at infinity while introducing a correction to the metric near the 4-brane. Note that the parameterization~\eqref{bm3} is only a subset of a more general form:
\begin{equation}
	\text{d}s_{D}^{2}
	 = 
	 A\,\text{d}R^{2}_{~}+B\,e^{f_{B}}_{~}\,\text{d}\Omega_{d-1}^{2}-C\,e^{f_{T}}_{~}\,\text{d}T^{2}_{~}+\,C\,e^{f_{s}}_{~}\,\text{d}\Sigma^{2}_{3}\,,
\end{equation}
where
\begin{subequations}
	\begin{eqnarray}
		f_{B}^{~}
		 &=& 
		\mathcal{B}_{0}^{(1)}|R-R_{0}^{~}|+\mathcal{B}_{0}^{(2)}(R-R_{0}^{~})^{2}_{~}+\ldots\,,\\
		f_{T}^{~}
		 &=& 
		\mathcal{C}_{T0}^{(1)}|R-R_{0}^{~}|+\mathcal{C}_{T0}^{(2)}(R-R_{0}^{~})^{2}_{~}+\ldots\,,\\
		f_{s}^{~}
		 &=& 
		\mathcal{C}_{s0}^{(1)}|R-R_{0}^{~}|+\mathcal{C}_{s0}^{(2)}(R-R_{0}^{~})^{2}_{~}+\ldots\,.
	\end{eqnarray}
\end{subequations}
Apparently, the more terms we reserve in the exponents, the more precise the parameterization becomes. However, this improvement in precision comes at the expense of increasing computational complexity.

The metric perturbation is generally time-dependent, since its source can evolute with the bulk time. In addition, the evolution of the matter on the 4-brane could also trigger the motion of the 4-brane in the extra space. Unlike the static 4-brane, a dynamical 4-brane can contribute extra corrections to the background spacetime. We thus employed the parameterization~\eqref{bm4} of the perturbed metric by a time-dependent generalization of~\eqref{bm3}. Similarly to the previous case, the parameters therein were derived from the singular part of the linearized perturbation equations. It does not imply that the parameterization~\eqref{bm4} only refers to a linearized perturbed metric. Indeed, the linearized perturbation only corresponds to the leading-order approximation of~\eqref{bm4} with respect to $|R-\mathcal{R}|$. Therefore, the parameters in the full solutions also satisfy the linearized perturbation equations. Furthermore, we proved that under the low-energy approximation, the metric perturbation is stable and remains small during the motion of the 4-brane. In fact, this nature is reserved by the nonlinear part of the perturbation. It can be verified that, with a time-dependent generalization of the linearized perturbed metric, $\mathcal{B}^{(1)}_{~}(R)\sim R\,|R-R_{0}^{~}|$ for instant, it becomes $\mathcal{B}^{(1)}_{~}(R,T)\sim \mathcal{R}\,|R-\mathcal{R}|$. So the linearized perturbations will monotonically increase to $\mathcal{B}^{(1)}_{~}(R,T)/B\sim1$ with the expansion of the 4-brane through $\mathcal{R}(T)=a(t)$. The linearization of the system would undergo an instability. Therefore a perturbative dynamical braneworld model with only the linearized perturbation would be unstable.

In addition to the stability of the dynamical braneworld model, the cosmology on the 4-brane is also an important topic. In our model, the brane cosmology is related to the equations that govern the dynamics of the 4-brane through the mapping between two metrics defined on the 4-brane, i.e., the induced metric~\eqref{im1} and the FLRW metric~\eqref{FLRW1}. The relation~\eqref{bcr2} does not alter the nature of spacetime. Thus the accelerating expansion of the 4-brane clearly indicates the presence of  dynamics for the 4-brane in the bulk. Although a delicate cancellation of the dynamics terms in the field equations might assist in achieving the accelerated expansion of the universe on the 4-brane fixed in the (static) bulk, it would be so tricky that we would not expect it to happen. Assuming that extra dimensions give small corrections to the cosmology for the late-time universe, we introduced the parameterization~\eqref{p2} to each of the cosmological parameters in the field equations, and then derived the modified Friedmann equations. Unlike 4-dimensional theories, the curvature of spacetime in the context of cosmology is determined by the effective cosmological constant rather than the bare one. It was found that by balancing the bare cosmological constant and the bulk cosmological constant [see Eq.~\eqref{ft1}], the standard Firedmann equations with a vanishing effective cosmological constant could be recovered at leading order. Note that in this case, the parameters in the modified Friedmann equations could still be solved analytically. However, once the effective cosmological constant is nonvaninshing, the calculation becomes significantly more complicated. Therefore we introduced the low-energy approximation to simplify the field equations. In this case, the fine-tuning~\eqref{bcbc1} is required, and the parameters in the modified Friedmann equations can only be solved numerically. We found that when the $\text{AdS}_{4}^{~}$ brane is embedded in the bulk, there are two branches of solutions for the effective cosmological constant. Our numerical results also support the existence of the universe with a positive effective cosmological constant, which is consistent with the dS expansion of our universe. In conclusion, our model allows for the presence of a 4-brane with a vanishing effective cosmological constant, an $\text{AdS}_{4}^{~}$ brane, and more importantly, a $\text{dS}_{4}^{~}$ brane in the bulk.

\section*{Acknowledgements}

We thank Yu-Xiao Liu and Wu-Zhong Guo for useful discussions. This work is supported by the National Key Research and Development Program of China under Grant No. 2020YFC2201504 and the National Natural Science Foundation of China under Grants Nos. 12247142 and 12047564.

\end{document}